\documentclass[sigconf]{acmart}

\usepackage{amsthm}
\usepackage{amsmath}
\usepackage{graphicx}
\usepackage{multirow}
\usepackage{tabularx}
\usepackage{subfigure}
\usepackage{pifont}
\usepackage{color}

\AtBeginDocument{%
  }

\setcopyright{acmlicensed}
\copyrightyear{2026}
\acmYear{2026}
\acmDOI{XXXXXXX.XXXXXXX}
\acmConference[The ACM Web Conference '26]{The Web Conference 2026}{April 13--17,
  2026}{Dubai, United Arab Emirates}
\acmISBN{978-1-4503-XXXX-X/2018/06}

\begin{document}

\title{Bridging Semantic Understanding and Popularity Bias with LLMs}

\author{Renqiang Luo}
\affiliation{%
    \institution{Jilin University}
    \city{Changchun}
    \country{China}
}
\email{lrenqiang@outlook.com}

\author{Dong Zhang}
\affiliation{%
    \institution{Dalian University of Technology}
    \city{Dalian}
    \country{China}
    }
\email{dongzhang1222@outlook.com}

\author{Yupeng Gao}
\affiliation{%
    \institution{Dalian University of Technology}
    \city{Dalian}
    \country{China}
}
\email{commer@mail.dlut.edu.cn}

\author{Wen Shi}
\affiliation{%
    \institution{Jilin University}
    \city{Changchun}
    \country{China}
}
\email{shiwen24@mails.jlu.edu.cn}

\author{Mingliang Hou}
\affiliation{%
    \institution{Jinan University \& TAL Education Group}
    \city{Guangzhou}
    \country{China}
}
\email{teemohold@outlook.com}

\author{Jiaying Liu}
\affiliation{%
    \institution{Dalian University of Technology}
    \city{Dalian}
    \country{China}
}
\email{jiayingliu@dlut.edu.cn}

\author{Zhe Wang}
\affiliation{%
    \institution{Jilin University}
    \city{Changchun}
    \country{China}
}
\email{wz2000@jlu.edu.cn}

\author{Shuo Yu*}
\authornote{Corresponding author.}
\affiliation{%
    \institution{Dalian University of Technology}
    \city{Dalian}
    \country{China}
}
\email{shuo.yu@ieee.org}

\renewcommand{\shortauthors}{Luo et al.}

\begin{abstract}
Semantic understanding of popularity bias is a crucial yet underexplored challenge in recommender systems, where popular items are often favored at the expense of niche content. 
Most existing debiasing methods treat the semantic understanding of popularity bias as a matter of diversity enhancement or long-tail coverage, neglecting the deeper semantic layer that embodies the causal origins of the bias itself.
Consequently, such shallow interpretations limit both their debiasing effectiveness and recommendation accuracy.
In this paper, we propose FairLRM, a novel framework that bridges the gap in the semantic understanding of popularity bias with Recommendation via Large Language Model (RecLLM). 
FairLRM decomposes popularity bias into item-side and user-side components, using structured instruction-based prompts to enhance the model’s comprehension of both global item distributions and individual user preferences. 
Unlike traditional methods that rely on surface-level features such as "diversity" or "debiasing", FairLRM improves the model's ability to semantically interpret and address the underlying bias. 
Through empirical evaluation, we show that FairLRM significantly enhances both fairness and recommendation accuracy, providing a more semantically aware and trustworthy approach to enhance the semantic understanding of popularity bias.
The implementation is available at~\url{https://github.com/LuoRenqiang/FairLRM}.
\end{abstract}




\maketitle

\section{Introduction}
\par As recommender systems strive to become more trustworthy, enhancing semantic understanding of core echical concepts, such as popularity bias, is essential for advancing cognitive capabilities of web applications~\cite{zhang2025mitigating, gao2025sprec}. 
Popularity bias occurs when popular items are recommended too frequently, while other products that may deserve more attention are ignored~\cite{abdollahpouri2019the}. 
This surface-level issue, however, masks a more complex interaction between user preferences, item rankings, and recommendation algorithms~\cite{tang2024text,ZhangYCLLXZ25}.
In this context, semantic understanding refers to a model's ability to grasp the deeper implications of popularity bias, rather than merely recognizing its surface-level features~\cite{xu2025comprehensive}.
In other words, a comprehensive semantic understanding of popularity bias requires recognizing not only the observable imbalance in recommendations but also the underlying causal mechanisms that give rise to it.

\begin{figure}[t]
    \centering
    \includegraphics[width=0.4\textwidth]{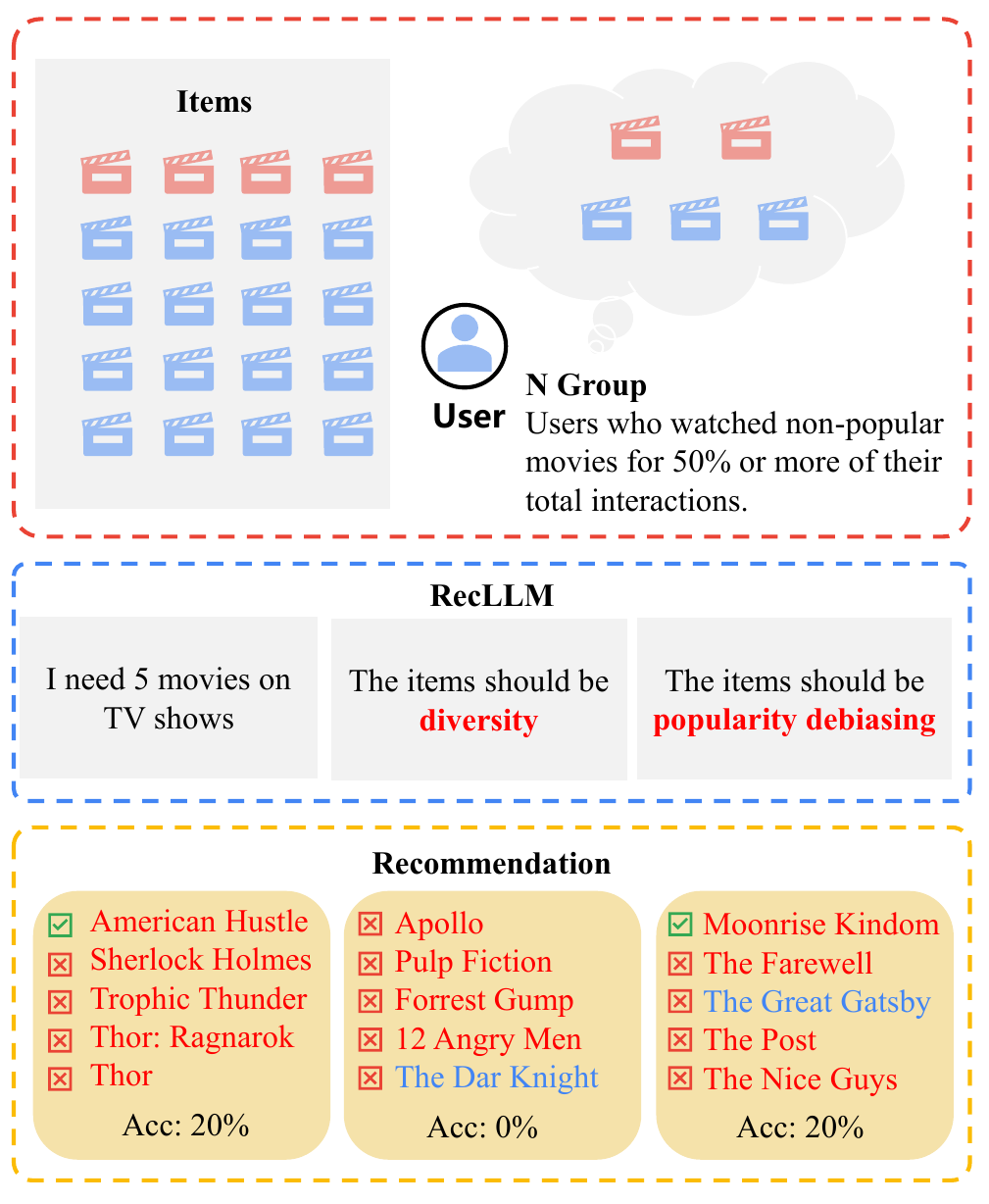}
    \caption{Even with prompts including "diversity" and "debiasing", RecLLM recommendations remain heavily concentrated on popular items.
    Red (symbols and titles) denotes popular items, and blue denotes the niche items.
    The persistent overemphasis on popular content limits relevance for users with niche or diverse preferences, indicating a lack of deep semantic understanding of popularity bias.}
    \label{fig:back}
\end{figure}

\par Developing a deeper semantic understanding of popularity bias is key to addressing challenges such as the long-tail problem~\cite{zhang2025deconfounding}, product homogeneity~\cite{lin2025how}, and personalized user satisfaction~\cite{silva2021exploiting, zhang2025survey}.
Existing methods, such as PAAC~\cite{cai2024popularity}, primarily focus on surface-level mitigation by increasing recommendation diversity or boosting exposure to non-popular items.
Although terms like “diversity” and “debiasing” are frequently used in prompts, our investigation with Large Language Models (LLMs) reveals that such cues remain shallow representations of popularity bias~\cite{klimashevskaia2024a, xiang2025use}.
They fail to encode the causal semantics that link user behavior, item features, and system design to the emergence of bias.

\par Our empirical observations confirm this limitation: RecLLMs not only inherit popularity bias but also amplify it, even when explicitly prompted with “diversity” or “debiasing” instructions.
As shown in Figure~\ref{fig:back}, LLM-generated recommendations remain dominated by popular items, leading to clear performance disparities among users with niche preferences (see Section~\ref{sec:eo} for details).
Even when prompted to increase diversity, RecLLMs simply substitute different popular items, revealing that their semantic understanding remains result-oriented, without recognizing the underlying causal factors, such as user–item interaction imbalance, that drive the bias~\cite{ortega2024evaluating}.

\par Motivated by this gap, this paper investigates the semantic understanding of popularity bias in RecLLMs and addresses two central challenges:
(1) Causal understanding in semantics: Existing approaches often overlook the causes of popularity bias, focusing solely on item-side distribution, which limits their ability to capture the full semantics of the problem.
(2) Ethical comprehension in LLMs: How can LLMs be guided to internalize ethical concepts like fairness and bias, rather than relying on shallow lexical cues such as “diversity”?
To explore these questions, we conduct an empirical evaluation on multiple datasets, testing different prompt designs aimed at improving item-side fairness and validating the necessity of causal, dual-side understanding.
We further define a text-to-item matching pipeline that ensures normalization, catalog alignment, and robust handling of out-of-catalog items.
Two key metrics—Long-tail Coverage (LtC)~\cite{abdollahpouri2018popularity} and Mean Recommendation Match Contribution (MRMC)~\cite{silva2021exploiting}—are employed to measure item-side and user-side fairness, respectively.
These findings collectively demonstrate that incorporating both causal and dual-side semantics is essential for a deeper understanding of popularity bias.

\par To bridge the gap between semantic understanding and popularity bias, we propose FairLRM, a semantic approach to understanding popularity bias.
FairLRM explicitly decomposes popularity bias into item-side and user-side components, mitigating both through instruction-based prompting. 
Unlike previous methods that rely on vague cues like "diversity" or "debiasing", FairLRM conditions the LLM on structured popularity signals, enabling it to better account for both global item distributions and individual user preferences. 
This approach results in significant improvements in both fairness and recommendation accuracy, providing a practical path toward more trustworthy semantic recommender systems.

\par The main contributions of this work are summarized as follows:

\begin{itemize}
 \item \textbf{Enhancing Semantic Understanding of Popularity Bias.} 
 We identify the shortcomings in how current methods understand and address popularity bias, particularly in the context of RecLLMs, and demonstrate the need for a deeper semantic understanding of this issue.
 \item \textbf{Structuring and Quantifying Popularity Debiasing Concepts.} 
 By breaking down the concept of popularity debiasing and expressing it in a structured and data-driven manner, we develop a semantic framework that is more suitable for model comprehension.
 \item \textbf{Improving Popularity Debiasing and Recommendation Performance in RecLLM.} 
 Building on this semantic framework, our RecLLM model achieves improved popularity debiasing and enhanced recommendation performance.
\end{itemize}

\section{Related Work}

\subsection{Semantic Understanding for Recommender Systems}
\par As web applications continue to evolve, achieving a deeper semantic understanding of complex definitions is crucial for enhancing the cognitive capabilities of models~\cite{zhang2024semantic, zhang2025can,YangZ0XZSL025}.
Bhattacharya and Pandey~\cite{bhattacharya2024developing} developed an agricultural ontology for extracting relationships from text using Natural Language Processing, aiming to improve semantic understanding. 
Wang et al.~\cite{wang2024research} proposed a deep learning model based on the Transformer architecture, integrating a self-attention mechanism to capture long-distance dependencies in text for improved semantic understanding tasks.

\par Several researchers have focused specifically on semantic understanding for recommender systems~\cite{liu2025understanding, zhang2025rethinking,WanyanHM0C25,Ma0RHC25}. 
Shang et al.~\cite{shang2024personalized} used LLMs to integrate semantic understanding with user preferences, combining a fine-tuned Roberta semantic analysis model with a multi-modal user preference extraction mechanism. 
Similarly, Wan et al.~\cite{wan2024larr} leveraged LLMs for semantic understanding by incorporating real-time scene information into recommender systems, enhancing RecLLM efficiency without requiring the model to process entire real-time scene texts directly.

\par Despite these advancements, a significant gap remains in the current understanding of semantics: the lack of explicit attention to ethical definitions, such as popularity bias~\cite{zhang2024logical}. 
While LLMs excel at generating semantically relevant recommendations, the training data often reflects existing popularity distributions, leading to the overexposure of popular items and the neglect of niche content~\cite{nie2024a}. 
This imbalance can result in unfair outcomes for users with diverse preferences, negatively impacting both fairness and user satisfaction.
Recent studies have started exploring fairness in RecLLM through fine-tuning and related techniques~\cite{jiang2024item, hua2024up5}, primarily focusing on group fairness (e.g., ensuring parity across demographic groups). 
However, the semantic understanding of popularity bias, particularly its impact on the interaction between items and users, remains largely unaddressed, leaving a significant gap in the current research landscape.

\subsection{Popularity Bias}
\par Popularity bias is a well-documented and pervasive issue in recommender systems, where popular items are disproportionately recommended, often at the expense of less popular, yet potentially relevant, long-tail items (or niche items)~\cite{daniil2024reproducing, luo2024algorithmic}. 
This bias can arise from multiple factors, including data sparsity for unpopular items, collaborative filtering mechanisms that naturally favor frequently interacted items, and optimization objectives that prioritize overall accuracy over diversity or fairness~\cite{chang2024cluster, luo2024fairgt}. 
The consequences of popularity bias are far-reaching: it can reinforce existing popularity distributions, limit user exposure to novel or niche content, and ultimately reduce both the utility and serendipity of recommendations~\cite{mullner2024the}. 
Additionally, the "rich-get-richer" phenomenon created by popularity bias can stifle innovation and hinder the visibility of emerging or new content~\cite{ning2024debiasing}.

\par Research has focused on mitigating popularity bias in traditional recommender systems~\cite{chen2024graph, ungruh2024putting}. 
Some approaches explicitly diversify recommendation lists by promoting long-tail items, while others modify learning objectives to penalize the recommendation of popular items~\cite{huang2024going}. 
These methods have shown varying levels of success in mitigating bias in traditional settings.
However, the direct application of these conventional debiasing techniques to the RecLLM paradigm is not straightforward due to the distinct architectural and operational characteristics of LLMs. 
LLMs' black-box nature and their reliance on vast amounts of pre-trained data make it difficult to apply traditional debiasing algorithms directly~\cite{dai2025unifying,ZhaoZ0R0LF025,ren2025causal}. 
Moreover, in the emerging field of RecLLM, the study of popularity bias and its mitigation remains underdeveloped. 
Existing attempts to address this issue in LLMs often rely on basic data partitioning or simplistic re-ranking strategies based on item popularity, which fall short of comprehensively tackling the problem~\cite{ortega2024evaluating, luo2025fairgp}.

\par A significant gap in current research lies in the lack of effective and actionable methods for debiasing popularity bias in LLMs, especially through the lens of semantic understanding. 

\begin{figure*}[t]
    \centering
    \includegraphics[width=0.8\textwidth]{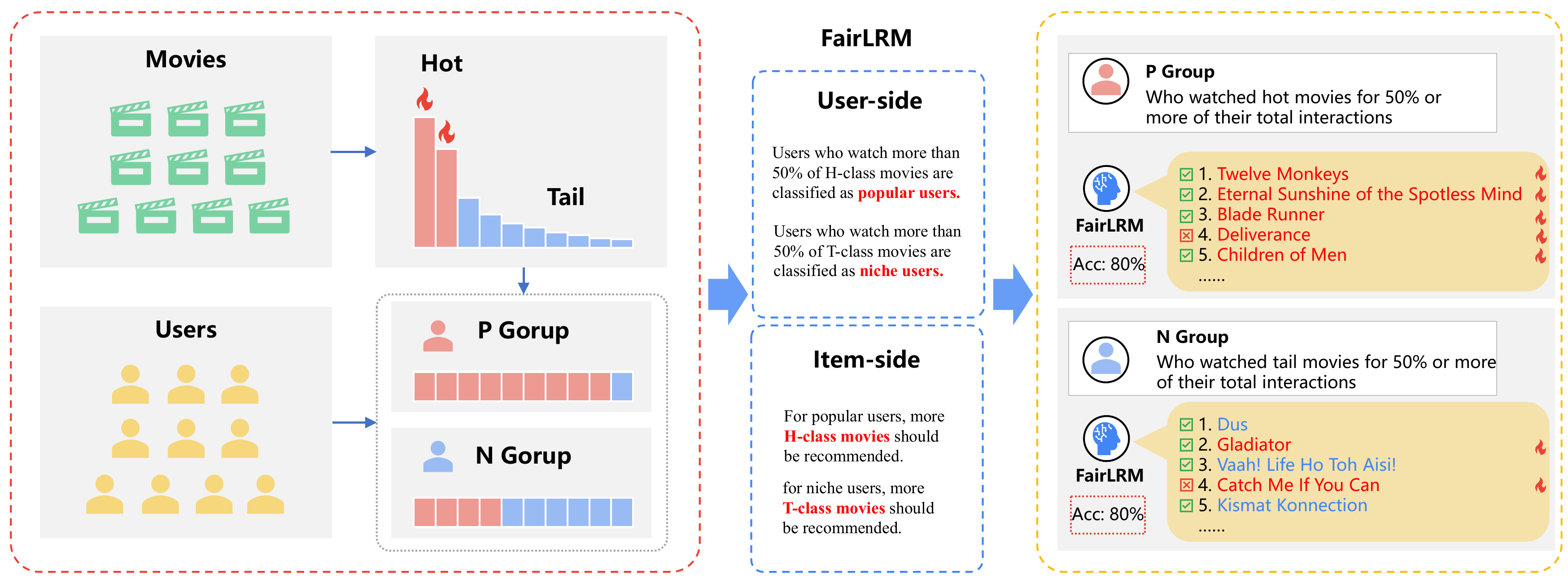}
    \caption{This illustrates how FairLRM's semantic understanding of popularity bias enhances recommendation performance, particularly for users with niche preferences.
    Red items are popular, while blue represents niche items (or long-tail items).}
    \label{fig:illustration}
\end{figure*}

\section{Prelimiaries}
\subsection{Notations and Metrics}
\par Unless otherwise specified, this study employs consistent notation. 
Sets are represented by copperplate uppercase letters (e.g., $\mathcal{A}$), matrices by bold uppercase letters (e.g., $\mathbf{A}$), and vectors by bold lower-case letters (e.g., $\mathbf{a}$).
The cardinality of a set $\mathcal{A}$ is denoted by $|\mathcal{A}|$.
The set of all users is defined as $\mathcal{U}$, with users in the test set specifically denoted as $\mathcal{U}_t$.
The complete set of items is represented by $\mathcal{I}$.
Within the item set, popular items are designated as $\mathcal{I}_p$, and niche items are denoted as $\mathcal{I}_n$.
The recommendation list generated for a user $u_i$ is represented by $\mathbf{L}(u_i)$, and $\mathbf{L}@k$ represents the top-$k$ items in this list.

\par This section outlines the metrics used to evaluate both accuracy and fairness of recommendation systems.
The metrics are chosen to comprehensively assess performance while explicitly capturing the extent of popularity bias.
LtC measures the diversity of long-tail items exposed to users, which is crucial in mitigating popularity bias~\cite{abdollahpouri2018popularity}. 
Consider a recommender system that repeatedly reccommends the same small set of long-tail items $\mathcal{I}_l$ to every user $u_i$. 
Such a system may perform well on metrics like Recommendation Popularity (RP) and Average Percentage of Long-tail items (APL), but fails to expose users to a wide range of long-tail items. 
LtC solves this issue by measuring the fraction of unique long-tail items across all recommendation lists $\mathbf{L}(u_i)$ for users in the test set $\mathcal{U}_t$, relative to the total number of long-tail items available:
\begin{equation}
    \text{LtC} = \frac{|(\cup_{u_i \in \mathcal{U}_t} \mathbf{L}({u_i})) \cap \mathcal{I}_n}{|\mathcal{I}_n|},
\end{equation}
where a higher LtC value indicates broader exposure of long-tail items and hence reduced popularity bias.
LtC shows the semantic understanding of popularity bias in item-side.

\par Mean Rank Miscalibration (MRMC) provides a normalized fairness measure within $[0, 1] \in \mathbb{R}$, enabling comparisons across different fairness criteria~\cite{silva2021exploiting}.
Formally, let $p$ denote the target distribution (e.g., global head/tail share or user-specific target), and let $q$ denote the empirical distribution constructed from recommendations at cutoffs $k$ over popularity bins.
The calculation of MRMC is structured as follows:
\begin{equation}
    \begin{aligned}
        & \text{MC}(p, q)=\frac{F(p, q)}{F(p, q(\{ \}))},
        \quad \text{RMC}(u_i)=\frac{\sum_{k=1}^N \text{MC}\left(p, q\left(\mathbf{L} @ k\right)\right)}{N}, \\
        & \text{MRMC} =\frac{\sum_{u_i \in U} \text{RMC}(u_i)}{|\mathcal{U}|},
    \end{aligned}
\end{equation}
where $F(\cdot,\cdot)$ is a fairness divergence function (e.g., KL divergence, Hellinger distance, or Chi-squared divergence).
The denominator $F(\cdot, \cdot(\{\}))$ corresponds to a void recommendation list, representing the worst-case divergence.
Thus, MC$(p,q)$ is a normalized miscalibration value, RMC$(u_i)$ averages this over top-$N$ positions for a given user, and MRMC aggregates over all users.
Smaller MRMC values indicate lower popularity bias and better fairness.
MRMC shows the semantic understanding of popularity bias in user-side.

\par MRR evaluates recommendation accuracy by measuring how highly the first relevant item is ranked.
If $rank_{u_i}$ denotes the position of the first relevant item for user $u_i$, MRR is defined as:
\begin{equation}
    \text{MRR} = \frac{1}{|\mathcal{U}|}\sum_{u_i \in \mathcal{U}}\frac{1}{rank_{u_i}},
\end{equation}
where higher MRR values indicate that relevant items are placed earlier in recommendation lists.

\par F1 combines precision@k and recall@k into a single harmonic mean, providing a balanced measure of recommendation quality.
For a user $u_i$, let $\mathbf{L}@k$ denote the top-$k$ recommended items and $\mathcal{I}(u_i)$ the ground-truth relevant items.
Then:
\begin{equation}
    \begin{aligned}
        &\text{Precision}@k(u_i) = \frac{|\mathbf{L}@k \cap \mathcal{I}(u_i)|}{k},\\
        &\text{Recall}@k(u_i) = \frac{|\mathbf{L}@k \cap \mathcal{I}(u_i)|}{|\mathcal{I}(u_i)|}.
    \end{aligned}
\end{equation}

\par The F1-score is given by:
\begin{equation}
    \text{F1}@k(u_i) = \frac{2 \cdot \text{Precision}@k(u_i) \cdot \text{Recall}@k(u_i)}{\text{Precision}@k(u_i) + \text{Recall}@k(u_i)}.
\end{equation}

\par Finally, F1@k is computed by averaging across all users:
\begin{equation}
    \text{F1}@k = \frac{1}{|\mathcal{U}|} \sum_{u_i \in \mathcal{U}} \text{F1}@k(u_i).
\end{equation}

\subsection{Item and User Popularity Categorization}
\par To inform the debiasing methodology, we establish a systematic categorization of both items and users based on popularity.
This process begins with item classification.
Leveraging the Pareto principle (i.e., the $20/80$ rule), items are sorted by their interaction counts~\cite{bender1981mathematical}.
Empirical evidence from long-tail distributions shows that a small fraction of items typically accounts for the majority of interactions.
Accordingly, the top $20\%$ of items with the highest interaction counts are designated as \textbf{P} (popular items or hot items), while the remaining $80\%$ are classified as \textbf{N} (niche items or tail items).
This $2/8$ division is consistent with the statistical characteristics of real-world recommender data, where item consumption naturally follows a long-tail distribution~\cite{zhao2025hierachical}.
The distinction between popular and niche items thus provides a principled basis for analyzing item popularity and guiding debiasing strategies, enabling the model to differentiate between widely consumed and less frequently accessed content.

\par Building upon this item categorization, users in the training set are grouped according to their historical engagement patterns, thereby capturing their popularity preference.
This categorization is conducted carefully to avoid any data leakage from the test set, thereby ensuring evaluation integrity.
Users are grouped as follows:
\begin{itemize}
    \item \textbf{P Group (Popularity-biased Users):} Users whose interactions with popular movies account for at least $50\%$ of their total consumption.
    This group is typically overserved by recommendation systems, often receiving redundant popular content that may overlook their more diverse interests.
    \item \textbf{N Group (Niche-biased Users):} Users whose interactions with popular items account for at most $50\%$ of their total consumption.
    These users are especially vulnerable to popularity bias, as their preferred niche items are systematically underrepresented. 
    Consequently, they often receive suboptimal recommendations that fail to reflect their true preferences.
\end{itemize}

\par This detailed user segmentation provides a structured foundation for developing a dual-side debiasing approach.
By explicitly distinguishing users who predominantly prefer popular items, those who prefer niche content, and those with diverse interests, the system can better tailor its recommendations.

\section{Methodology}
\par Popularity bias arises when recommendation models disproportionately favor items with high interaction counts (popular items), while overlooking long-tail or niche items that better align with certain user groups~\cite{sipio2025addressing}. 
This bias not only reduces exposure to diverse content but also creates unfair disadvantages for users whose tastes lie outside the mainstream.
Thus, semantic understanding of popularity bias is critical from both the item-side and user-side perspectives.
This section presents our methodology for achieving a semantic understanding of popularity bias with RecLLM, focusing on prompt optimization through semantic guidance. 
Our approach revolves around two central questions: 
(1) Are traditional debiasing prompts, such as "diversity" or "debiasing", effective when applied to RecLLM? 
(2) How can we enable LLMs to semantically understand and actively mitigate popularity bias, especially from a dual-side perspective. 
Figure~\ref{fig:illustration} provides an overview of the framework.

\subsection{Empirical Observations} \label{sec:eo}
\par We conduct experiments on the MovieLens-$20$M dataset, splitting it into a training set ($70$\%) and a test sets ($30$\%). 
Based on the number of interactions in the training set, we categorize the top $20$\% of movies as popular and the remaining $80$\% as niche. 
We then tested RecLLM prompts with three different strategies: vanilla RecLLM, adding "diversity" to the prompts, and incorporating "popularity debiasing" directly into the prompts.

\par The results, shown in Figure~\ref{fig:back}, reveal that while adding diversity slightly increases the recommendation of niche movies, it significantly compromise the model's performance, particularly in recommendation accuracy. 
On the other hand, incorporating popularity debiasing results in some increase in niche movie recommendations, but still fails to adequately meet the needs of N Group users, who require a sufficient number of niche items.

\par These findings suggest that "diversity" and "debiasing" are surface-level features in recommender systems and do not fully address the deeper semantic understanding required to mitigate popularity bias. 
This highlights the need for more advanced strategies that go beyond these surface-level adjustments.
The prompt details are as follows:
\textbf{Vanilla RecLLM}: "I need $10$ movies or TV shows."
\textbf{diversity}:"Please recommend a diverse list of $10$ movies."
\textbf{popularity debiasing}:"Please apply popularity debiasing: avoid recommending only popular movies; include long-tail movies when appropriate."

\subsection{Diversity Requirement in Prompts}
\par A natural starting point is to include explicit diversity requirements in the prompts given to the LLM. 
The goal is to test whether established debiasing methods, which typically encourage item diversity, can be effecitvely applied in the RecLLM context.
The prompt then reads:"Please recommend a diverse list of 10 movies, covering different genres and popularity levels."
This ensures the model understands the diversity of items from an item-side perspective.

\par While this approach increases the apparent variety of recommendations, the results show that LLMs largely achieve “diversity” by staying within the popular items. 
Instead of recommending niche items, the model tends to suggest multiple popular movies from different genres. 
As a result, the underlying popularity bias remains unresolved. 
In addition, precision often declines as the variety of recommended items does not align with user preferences.

\par This finding demonstrates a limitation of simply applying generic diversity constraints: the LLM interprets “diverse” semantically but defaults to popularity when filling the requirement. 
Without a deeper semantic grounding in what constitutes popularity bias, the model cannot effectively distinguish “true diversity” from superficial variation within popular items. 
This highlights the need for explicit bias grounding within the debiasing strategy.

\subsection{Dual-side Popularity Debiasing}
\par To overcome these limitations, we shift the focus to a dual-side, semantically grounded approach. 
Instead of treating all users as identical, we classify them based on their historical interaction patterns with popular and niche items, ensuring no data leakage from the test set. 
This user categorization is then incorporated into the prompt, providing the LLM semantic context about user preference. 
With this guidance, the model can adjust its recommendations to better balance popular and niche content for different user groups.

\par Importantly, this strategy goes beyond simply “asking for diversity". 
It enables the LLM to understand the semantic gap between users preferring popular items and those seeking niche or mixed content. 
The model can then conceptualizes popularity bias: for P Group users, it avoids over-serving popular items; for N Group users, it ensures sufficient exposure to niche items.

\par Since the effectiveness of user classification may vary with different LLMs, we tested multiple thresholds for categorizing user bias:
\textbf{FairLRM ($55$)}: A moderate threshold, where users are considered as P Group, if at least $50$\% of their interactions involve popular items. 
\textbf{FairLRM ($82$)}: A stricter threshold, requiring $80$\% of interactions to involve popular items before a user is classified as P Group. 
This isolates users with a very strong popularity bias, allowing for targeted analysis.
By embedding these categorizations directly into the prompts, FairLRM enables LLMs to reason semantically about user preferences and popularity bias, rather than relying on superficial diversity signals. 
This method improves both fairness and personalization, addressing popularity bias more effectively than diversity-only constraints.
Subsequently, the user classification information is embedded explicitly within the prompt for each recommendation request.
The prompt states:

\textit{“The user segmentation rules are as follows: users who watch more than $50$\% of H-class movies are classified as popular users, those who watch more than $50$\% of T-class movies are classified as niche users, and the remaining are ordinary users. For popular users, more H-class movies should be recommended, while for niche users, more T-class movies should be recommended.”}

\par These segmentation details are adapted to the specific characteristics of the LLM in use, ensuring optimal comprehension and effective guidance.
The careful design transforms the LLM from a general text generator into a bias-aware recommender system.
By explicitly encoding a user’s preference for popular or niche content, the LLM learns that popularity bias originates from heterogeneous user preferences rather than solely item-level frequency.
This direct integration of user-side popularity information aims to enhance the LLM’s semantic understanding of bias, guiding it to produce recommendations that are both personalized and fair.
Unlike broad diversity requirements, which often neglect user-specific needs, this approach aligns debiasing with individual preference structures.
As a result, it is expected to improve fairness metrics while maintaining or even enhancing accuracy, enabling a more equitable distribution of recommendations across user groups.

\begin{figure*}[t]
    \centering
    \includegraphics[width=0.8\textwidth]{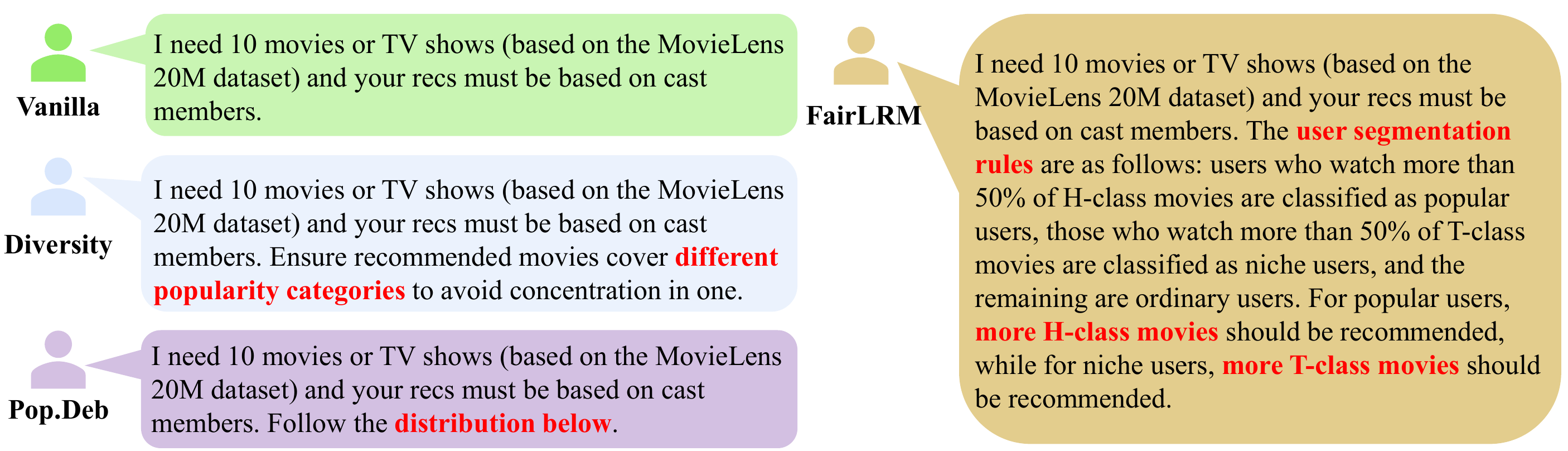}
    \caption{Prompts of baselines.}
    \label{fig:bas}
\end{figure*}

\begin{table}[t]
    \centering
    \footnotesize
    \caption{Statistical summary of the datasets.}
    \tabcolsep=0.3cm
    \renewcommand{\arraystretch}{0.85}
    \begin{tabular}{lcc}
        \toprule
        Datasets & MovieLens-$20$M & Goodbooks-$10$k \\
        \midrule
        The number of users         & $138,493$     & $53,271$  \\
        The number of items         & $270,000$     & $10,000$  \\
        The number of interaction   & $20,000,000$  & $5,972,476$ \\
        \textcolor{red}{P Group}/ \textcolor{blue}{N Group} ($55$)     & \textcolor{red}{$138,484$}/\textcolor{blue}{$9$} & \textcolor{red}{$45,599$}/\textcolor{blue}{$7,672$}\\
        \textcolor{red}{P Group}/ \textcolor{blue}{N Group} ($82$)     & \textcolor{red}{$138,019$}/\textcolor{blue}{$474$} & \textcolor{red}{$9,401$}/\textcolor{blue}{$43,880$}\\
        \bottomrule
    \end{tabular}
    \label{tab:results}
\end{table}

\section{Experiments}
\subsection{Datasets}
\par The MovieLens-$20$M dataset~\cite{harper2015movielens} validates the proposed methodology.
This widely recognized benchmark, collected from the MovieLens movie recommendation service, consists of $20$ million five-star ratings and free-text tags for more than $270,000$ movies by $138,493$ users.
Its large scale and diversity provide a rich source of real-world user–item interaction data, making it particularly suitable for evaluating LLM-based recommendation approaches under realistic conditions and for systematically observing popularity bias.

\par The Goodbooks-$10$k dataset~\cite{zajac2017goodbooks} contains $982,000$ ratings of $10,000$ books provided by $53,271$ users.
The dataset is publicly available through the FastML platform and includes ratings, book metadata, to-read tags, as well as user-generated tags and shelves.
Its rating distribution is centered around approximately $100$ ratings per user, with an average rating of about $4$.
Both the number of ratings per user and the average rating distribution follow patterns close to a multivariate normal distribution, making this dataset a complementary benchmark to MovieLens-$20$M for testing the robustness of the proposed method across different domains.

\par Dataset preprocessing follows a procedure across both benchmarks.
First, users with fewer than $30$ interactions are removed to ensure users with sufficient historical records are included, thereby mitigating the cold-start problem and providing more reliable preference signals for the model.
Second, the remaining interactions for each user are sorted in ascending order by timestamp.
Subsequently, the dataset is partitioned into training and testing subsets using a $70\%/30\%$ temporal split, with the most recent $30$\% of interactions reserved for evaluation.
This strategy captures the temporal dynamics of user preferences, prevents data leakage, and ensures that experimental results are realistic and robust.

\par Importantly, all user popularity categorizations (e.g., $50$\% or $80$\% thresholds distinguishing popular-leaning and niche-leaning users) are conducted exclusively on the training set.
This design guarantees that the evaluation process remains unaffected by these operations, thereby preserving the integrity of the test set and ensuring that reported results faithfully reflect the model’s generalization performance under unbiased conditions.

\begin{table}[t]
    \centering
    \footnotesize
    \caption{MovieLens-$20$M results with Qwen-max: higher MRR/F1 indicate better performance, while higher LtC and lower MRMC signify reduced popularity bias.}
    \tabcolsep=0.25cm
    \renewcommand{\arraystretch}{0.8}
    \begin{tabular}{lcccc}
        \toprule
        LLM & \multicolumn{4}{c}{\textbf{Qwen-max}} \\
        Metrics & LtC $\uparrow$ & MRMC $\downarrow$ & MRR@10 $\uparrow$ & F1@10 $\uparrow$ \\
        \midrule
         Vanilla & $0.013$ & $0.586$ & $0.078$ & $0.576$ \\
         Pop.Debiasing ($55$) & $0.046$ & $0.372$ & $0.403$ & $0.760$ \\
         Pop.Debiasing ($82$) & $0.050$ & $0.374$ & $0.402$ & $0.758$ \\
         Diversity ($55$) & $0.058$ & $0.384$ & $0.138$ & $0.751$ \\
         Diversity ($82$) & $0.048$ & $0.458$ & $0.122$ & $0.688$ \\
        \midrule
         FairLRM ($55$) & $\textcolor{red}{\mathbf{0.062}}$ & $\textcolor{red}{\mathbf{0.302}}$ & $\textcolor{red}{\mathbf{0.452}}$ & $\textcolor{red}{\mathbf{0.810}}$ \\
         FairLRM ($82$) & $0.058$ & $0.310$ & $0.448$ & $0.803$ \\
        \bottomrule
    \end{tabular}
    \label{tab:results_1}
\end{table}

\begin{table}[t]
    \centering
    \footnotesize
    \caption{MovieLens-$20$M results with Llama: higher MRR/F1 indicate better performance, while higher LtC and lower MRMC signify reduced popularity bias.}
    \tabcolsep=0.25cm
    \renewcommand{\arraystretch}{0.8}
    \begin{tabular}{lcccc}
        \toprule
        LLM & \multicolumn{4}{c}{\textbf{Llama}} \\
        Metrics & LtC $\uparrow$ & MRMC $\downarrow$ & MRR@10 $\uparrow$ & F1@10 $\uparrow$ \\
        \midrule
        Vanilla & $0.036$ & $0.370$ & $0.053$ & $0.686$ \\
        Pop.Debiasing ($55$) & $0.036$ & $0.365$ & $0.131$ & $0.770$ \\
        Pop.Debiasing ($82$) & $0.024$ & $0.363$ & $0.178$ & $0.779$ \\
        Diversity ($55$) & $0.027$ & $0.361$ & $0.144$ & $0.773$ \\
        Diversity ($82$) & $0.021$ & $0.379$ & $0.091$ & $0.760$ \\
        \midrule
        FairLRM ($55$) & $\textcolor{red}{\mathbf{0.048}}$ & $\textcolor{red}{\mathbf{0.331}}$ & $0.170$ & $\textcolor{red}{\mathbf{0.794}}$ \\
        FairLRM ($82$) & $0.029$ & $0.367$ & $\textcolor{red}{\mathbf{0.213}}$ & $0.780$ \\
        \bottomrule
    \end{tabular}
    \label{tab:results_2}
\end{table}

\subsection{Baselines}
\par To provide a comprehensive comparative analysis, the setup includes several baselines for FairLRM's performance measurement. 
These baselines encompass both the raw capabilities of LLMs and existing approaches to mitigate popularity bias and enhance diversity.
The fundamental comparison point is the vanilla LLM model, which represents the direct application of Qwen-max~\cite{bai2023qwen} and Llama~\cite{touvron2023llama} without any specific prompt engineering or explicit instructions for bias mitigation. 
This baseline reveals the inherent biases and recommendation characteristics when these models are used out-of-the-box for recommendation tasks, serving as a critical reference for measuring the impact of debiasing strategies. 
This helps to establish the initial performance landscape and the baseline level of popularity bias present in RecLLM, against which the effectiveness of debiasing methods can be quantified.

\par Beyond the vanilla models, two primary debiasing baselines are considered. 
Popularity Debiasing~\cite{ortega2024evaluating} (Pop.Debiasing) represents methods that leverage Zero-Shot LLMs to calibrate popularity bias and ensure fairness through carefully constructed prompts, similar to recent approaches in the literature. 
This baseline assesses the LLM's intrinsic capacity to debias when explicitly guided towards reducing popularity, typically by instructing the model to prioritize less popular items or to distribute recommendations more evenly across popularity strata.
The Diversity baseline serves as a control, where prompts are augmented with direct instructions to encourage the generation of diverse item lists. 
This helps evaluate whether a simple diversity mandate can effectively counter popularity bias without compromising recommendation quality, and also highlights the limitations of such generic approaches in complex fairness scenarios where diversity might be achieved at the cost of relevance.
The details of prompts are shown in Figure~\ref{fig:bas}.

\par Crucially, to investigate how user classification standards influence debiasing effectiveness, both the Pop. Debiasing and Diversity baselines are also evaluated under different user grouping criteria, denoted as ($55$) and ($82$). 
These suffixes represent the interaction thresholds that define user groups: 
($55$) indicatets users with at least $50\%$ of their interactions involving popular items, while ($82$) applies a stringent threshold of $80\%$ or more. 
This setup enables a fine-grained comparison of how varying user classification granularities affect the performance of these baseline strategies across different LLMs, offering insights into their robustness and sensitivity under different definitions of user popularity. 
Moreover, this evaluation underscores the pivotal role of user-side semantic understanding in interpreting and mitigating popularity bias.
Overall, this comprehensive baseline design ensures a fair and systematic evaluation of FairLRM against existing paradigms, emphasizing its strength as a dual-side approach to semantic debiasing.

\par Our experimental setup involves two LLMs:
For Qwen-max, we purchased and utilized its official API for deploying the experiments.
For Llama, we used the Llama-$7$B model, which was deployed locally on our server equipped with an NVIDIA L$40$ $48$GB GPU.

\subsection{Comparison Results}
\par This section presents a concise comparative analysis of FairLRM against baseline methods: naive popularity debiasing and diversity-focused approaches, across both the Qwen-max and Llama architectures.
Crucially, all metrics reported in this study, including those below, are evaluated at Top-$10$ (denoted as @$10$) to ensure a consistent comparison of results.
Table~\ref{tab:results_1}-\ref{tab:results_4} summarizes the comprehensive performance results, evaluating key metrics for fairness, accuracy, and ranking quality, such as LtC, MRMC, MRR, F1 Score. 
Higher values for LtC show better popularity debiasing in item-side and lower MRMC value signifies more effective popularity debiaisng in user-side. 
Higher MRR and F1 indicate superior performance.
The results underscore that the optimal user grouping standards, represented by the ($55$) and ($82$) thresholds, exhibit varying tendencies and impact across different LLM architectures. 
This highlights the importance of fine-grained user segmentation being tailored to specific model characteristics, a key insight demonstrated by the superior performance of FairLRM.

\begin{table}[t]
    \centering
    \footnotesize
    \caption{Goodbooks-$10$k results with Qwen-max: higher MRR/F1 indicate better performance, while higher LtC and lower MRMC signify reduced popularity bias.}
    \tabcolsep=0.25cm
    \renewcommand{\arraystretch}{0.8}
    \begin{tabular}{lcccc}
        \toprule
        LLM & \multicolumn{4}{c}{\textbf{Qwen-max}} \\
        Metrics & LtC $\uparrow$ & MRMC $\downarrow$ & MRR@10 $\uparrow$ & F1@10 $\uparrow$ \\
        \midrule
        Vanilla                 & $0.696$ & $0.274$ & $0.026$ & $0.064$ \\
        Pop.Debiasing ($55$)    & $0.810$ & $0.188$ & $0.023$ & $0.697$ \\
        Pop.Debiasing ($82$)    & $0.808$ & $0.189$ & $0.017$ & $0.682$ \\
        Diversity ($55$)        & $0.813$ & $0.198$ & $0.009$ & $0.690$ \\
        Diversity ($82$)        & $0.795$ & $0.191$ & $0.016$ & $0.705$ \\
        \midrule
        FairLRM ($55$)          & $\textcolor{red}{\mathbf{0.818}}$ & $0.182$ & $\textcolor{red}{\mathbf{0.035}}$ & $0.715$ \\
        FairLRM ($82$)          & $0.798$ & $\textcolor{red}{\mathbf{0.175}}$ & $0.029$ & $\textcolor{red}{\mathbf{0.745}}$ \\
        \bottomrule
    \end{tabular}
    \label{tab:results_3}
\end{table}

\begin{table}[t]
    \centering
    \footnotesize
    \caption{Goodbooks-$10$k results with Llama: higher MRR/F1 indicate better performance, while higher LtC and lower MRMC signify reduced popularity bias.}
    \tabcolsep=0.25cm
    \renewcommand{\arraystretch}{0.8}
    \begin{tabular}{lcccccc}
        \toprule
        LLM & \multicolumn{4}{c}{\textbf{Llama}} \\
        Metrics & LtC $\uparrow$ & MRMC $\downarrow$ & MRR@10 $\uparrow$ & F1@10 $\uparrow$\\
        \midrule
        Vanilla             & $0.319$ & $0.198$ & $0.341$ & $0.809$ \\
        Pop.Debiasing ($55$)  & $0.503$ & $0.183$ & $0.425$ & $0.816$ \\
        Pop.Debiasing ($82$)  & $0.449$ & $0.234$ & $0.425$ & $0.806$ \\
        Diversity ($55$)      & $0.628$ & $0.205$ & $0.144$ & $0.812$ \\
        Diversity ($82$)      & $0.513$ & $0.214$ & $0.226$ & $0.679$ \\
        \midrule
        FairLRM ($55$)        & $0.738$ & $0.168$ & $0.330$ & $0.820$ \\
        FairLRM ($82$)        & \textcolor{red}{$\mathbf{0.795}$} & \textcolor{red}{$\mathbf{0.166}$} & \textcolor{red}{$\mathbf{0.480}$} & \textcolor{red}{$\mathbf{0.822}$} \\
        \bottomrule  
    \end{tabular}
    \label{tab:results_4}
\end{table}

\begin{figure*}[t]
	\centering
	\subfigure {
		\begin{minipage}[b]{0.28\textwidth}
			\centering
			\includegraphics[width=1.0\textwidth]{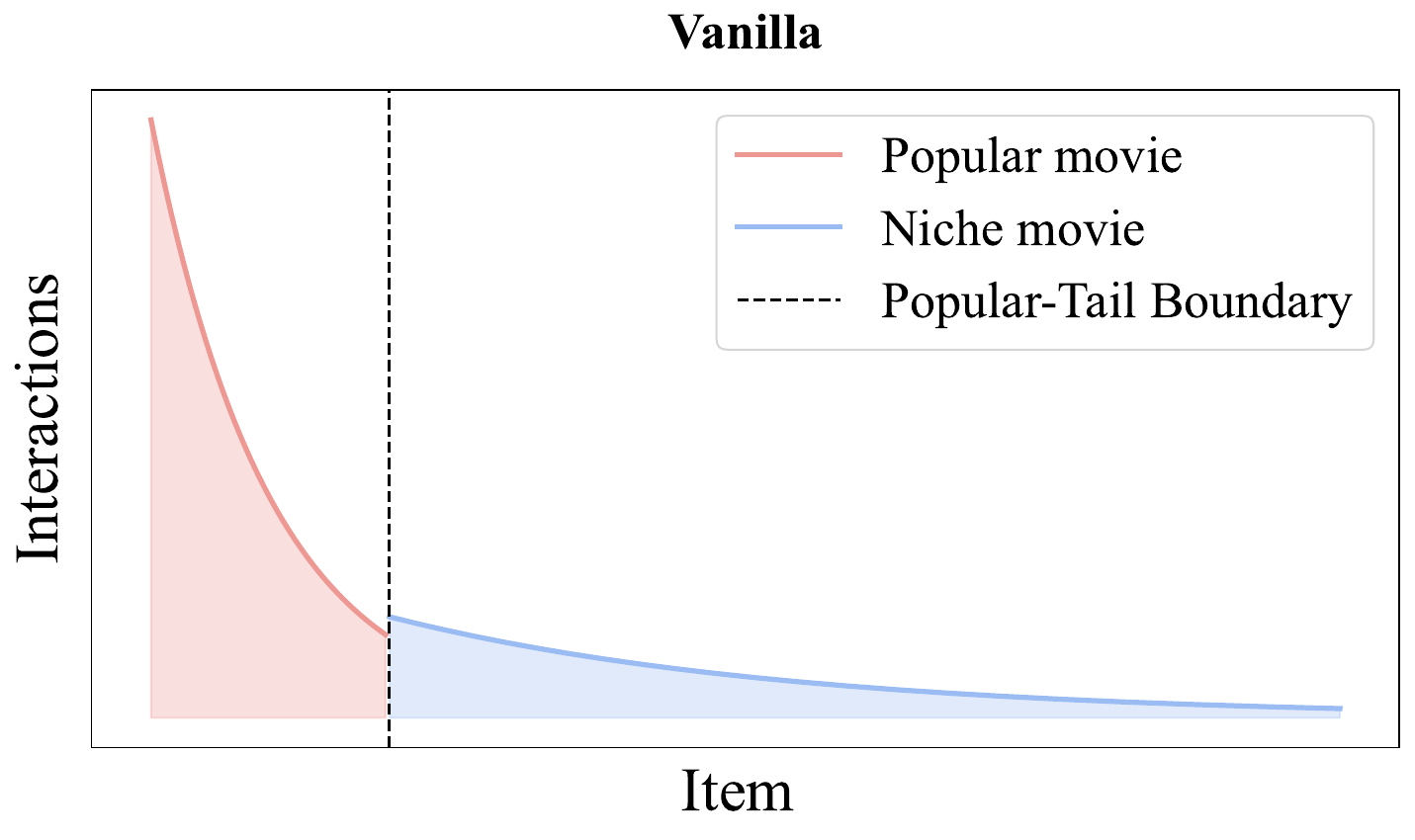}
		\end{minipage}
	}
    \subfigure {
		\begin{minipage}[b]{0.28\textwidth}
			\centering
			\includegraphics[width=1.0\textwidth]{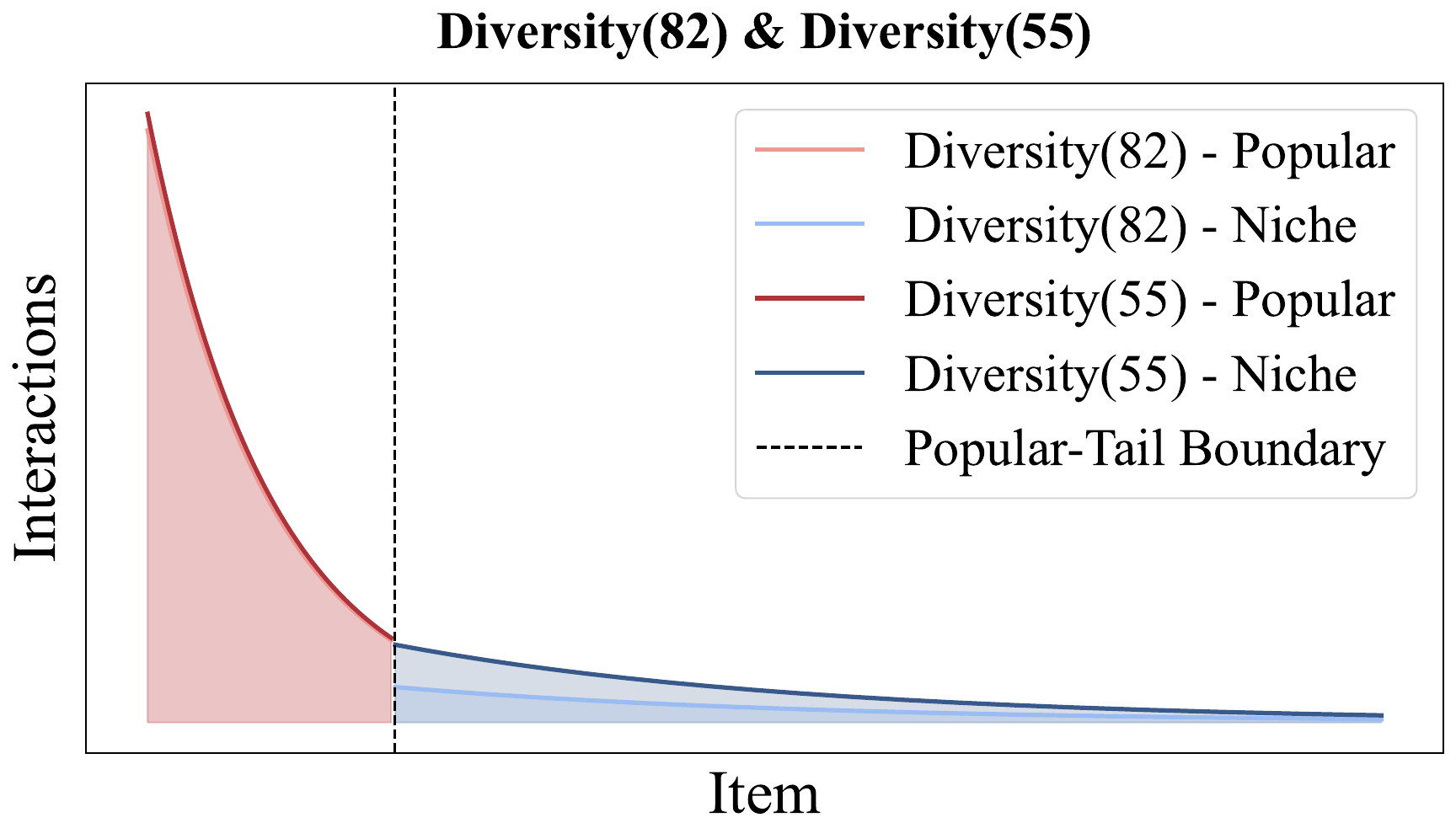}
		\end{minipage}
	}
    \subfigure {
		\begin{minipage}[b]{0.28\textwidth}
			\centering
			\includegraphics[width=1.0\textwidth]{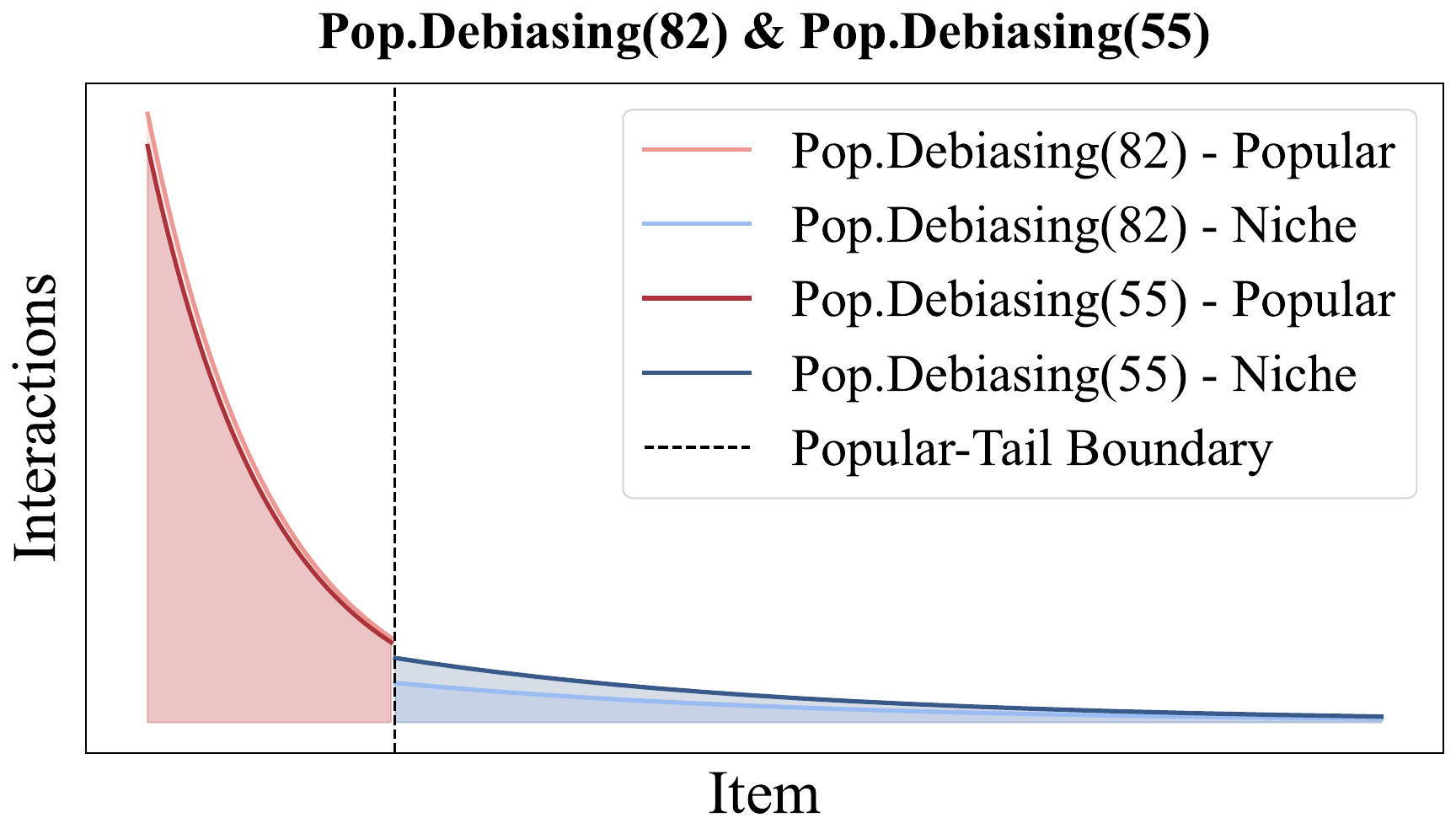}
		\end{minipage}
	}
	\subfigure {
		\begin{minipage}[b]{0.28\textwidth}
			\centering
			\includegraphics[width=1.0\textwidth]{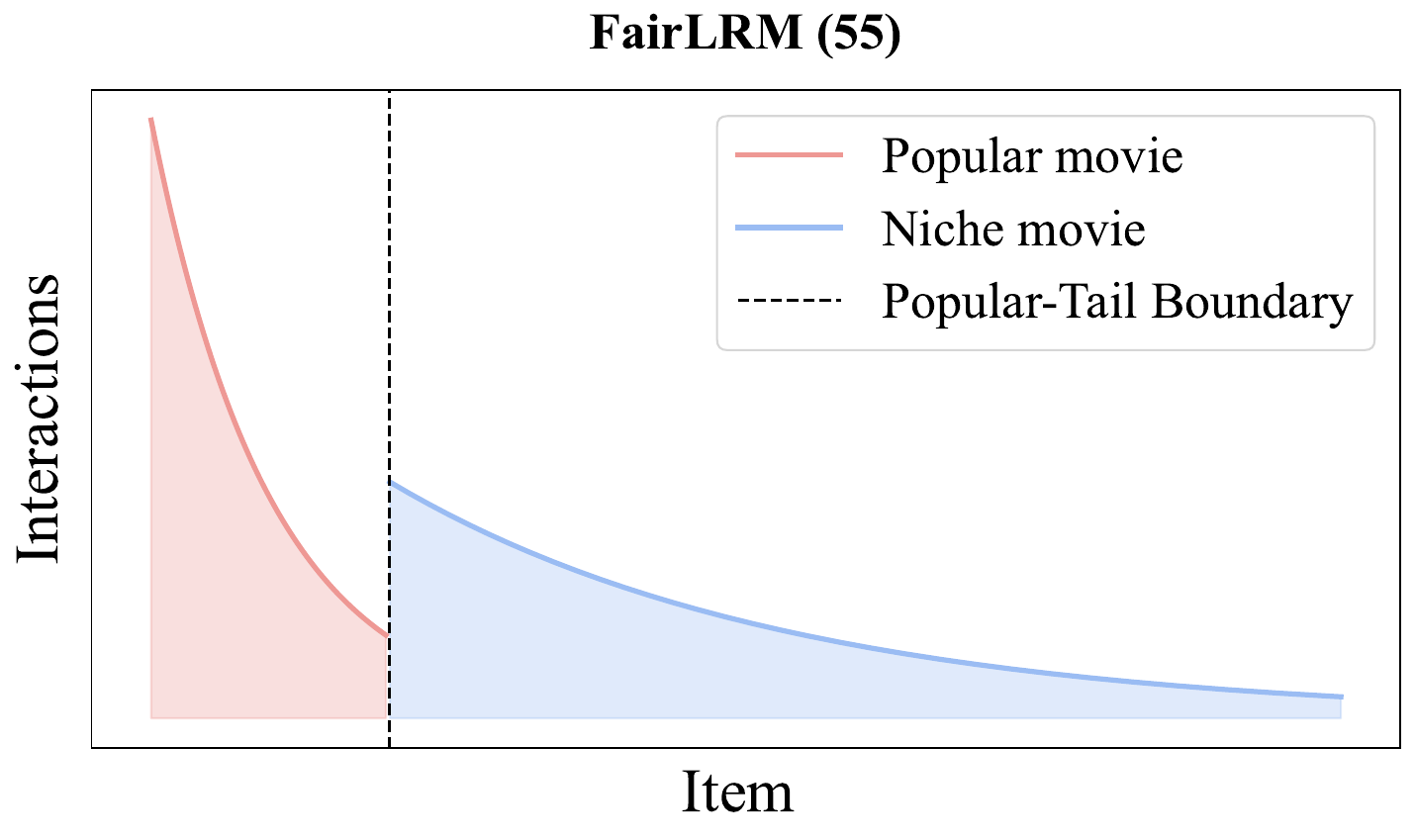}
		\end{minipage}
	}
    \subfigure {
		\begin{minipage}[b]{0.28\textwidth}
			\centering
			\includegraphics[width=1.0\textwidth]{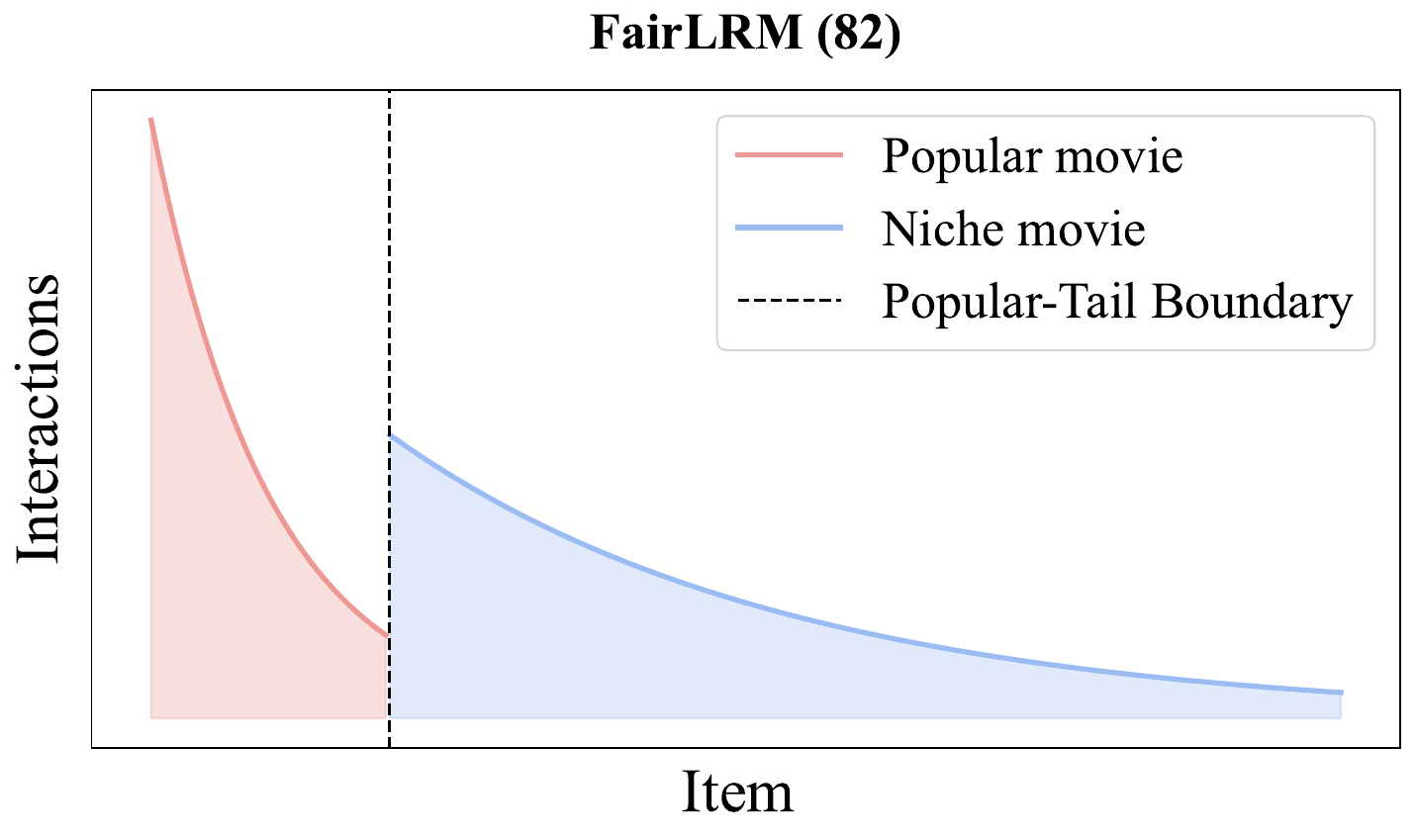}
		\end{minipage}
	}
  \caption{Distribution of recommended items across different methods: popular vs. niche items.}
  \label{fig:results}
\end{figure*}

\par The Pop.Debiasing baselines, which explicitly target popularity mitigation, exhibit inconsistent performance.
For Qwen-max across both datasets, they offer a moderate balance, with LtC improvements (e.g., $0.046/0.810$ for ($55$) and $0.050/0.808$ for ($82$), compared with Vanilla’s $0.013/0.696$).
However, in Goodbooks-$10$k, an aggressive debiasing strategy under the ($82$) setting drastically reduces MRR ($0.017$ vs. $0.026$ for Vanilla and $0.023$ for Pop.Debiasing ($55$)), indicating an impractical trade-off between fairness and accuracy.
These results highlight the difficulty of applying generic popularity-based debiasing methods without incorporating user-specific context.
In contrast, the Diversity baselines consistently show significant degradation in recommendation accuracy (MRR and F1) across both LLMs.
For instance, Qwen-max in Goodbooks-$10$k drops to MRR values of $0.009$ and $0.016$ for ($55$) and ($82$), respectively, compared with $0.026$ for Vanilla.
A similar pattern appears for Llama, where Diversity baselines underperform relative to both Pop. Debiasing and FairLRM.
This performance gap is largely attributed to the stronger user-side popularity bias among N group users in Goodbooks-$10$k, who tend to assign higher ratings, amplifying the influence of popular items.
These findings confirm that merely increasing diversity, without semantic understanding of user preferences, produces irrelevant recommendations that fail to counteract popularity bias, often sacrificing relevance for variety.

\par In comparison, FairLRM consistently achieves superior performance.
For Qwen-max, it attains the highest MRR ($0.425/0.035$) and the lowest MRMC ($0.302/0.182$), reflecting strong long-tail coverage and effective bias mitigation, while maintaining high accuracy (F1: $0.810/0.745$).
These results represent a clear improvement over both Vanilla and existing debiasing baselines, demonstrating FairLRM’s ability to balance fairness and accuracy simultaneously.
For Llama, FairLRM also achieves competitive MRMC ($0.331/0.166$) and robust F1 scores ($0.794/0.822$), outperforming its respective baselines.
Although FairLRM ($82$) on Llama shows higher MRR ($0.213$) than FairLRM ($55$) ($0.170$) in MovieLens-$20$M, its LtC is slightly lower, indicating that the user-grouping threshold can be tuned to adjust the balance between fairness metrics for different LLMs.
Overall, these results validate FairLRM’s strategy, which enhances fairness by semantically understanding user preference patterns, mitigating popularity bias without sacrificing recommendation quality.

\par Figure~\ref{fig:results} visualizes the distribution of recommended items across methods.
Items are color-coded, with red representing popular items and blue representing non-popular (tail) items, separated by a vertical black line.
Our method demonstrates a clear advantage in semantic understanding of popularity bias, as it recommends a significantly higher number of non-popular items.
Notably, many tail items are recommended more frequently than popular ones, underscoring FairLRM’s ability to mitigate popularity bias and promote genuinely diverse yet semantically relevant recommendations.

\begin{table}[t]
    \centering
    \footnotesize
    \caption{Performance comparison via diversity influence across MovieLens-$20$M (Transposed).
    Diversity represents prompt including diversity and item rank, and Item-rank represents item rank without diversity.}
    \tabcolsep=0.25cm
    \renewcommand{\arraystretch}{0.8}
    \begin{tabular}{lcccc}
        \toprule
        \textbf{Qwen-max} & LtC $\uparrow$ & MRMC $\downarrow$ & MRR@$10$ $\uparrow$ & F1@$10$ $\uparrow$ \\
        \midrule
        \textbf{Diversity} & $0.024$ & $0.399$ & $0.068$ & $0.748$ \\
        \textbf{Item-rank} & $0.045$ & $0.436$ & $0.117$ & $0.754$ \\
        \textbf{FairLRM} & \textcolor{red}{$\mathbf{0.062}$} & \textcolor{red}{$\mathbf{0.302}$} & \textcolor{red}{$\mathbf{0.452}$} & \textcolor{red}{$\mathbf{0.810}$} \\
        \midrule
        \textbf{Llama} & LtC $\uparrow$ & MRMC $\downarrow$ & MRR@$10$ $\uparrow$ & F1@$10$ $\uparrow$ \\
        \midrule
        \textbf{Diversity}  & $0.027$ & $0.361$ & $0.068$ & $0.773$ \\
        \textbf{Item-rank}  & $0.031$ & $0.340$ & $0.117$ & $0.784$ \\
        \textbf{FairLRM}    & \textcolor{red}{$\mathbf{0.048}$} & \textcolor{red}{$\mathbf{0.331}$} & \textcolor{red}{$\mathbf{0.170}$} & \textcolor{red}{$\mathbf{0.794}$} \\
        \bottomrule
    \end{tabular}
    \label{tab:results_diversity_transposed}
\end{table}

\subsection{Ablantion Study}
\par Further experiments specifically evaluated whether simply adding the term diversity in prompts, alongside explicit item-rank considerations, can help LLMs mitigate popularity bias as effectively as traditional recommender systems.
Table~\ref{tab:results_diversity_transposed} compares a Diversity baseline, which includes both diversity and item-rank instructions, with an "Item-rank" baseline that excludes explicit diversity.

\par The results demonstrate that, unlike traditional recommender systems where diversity constraints effectively mitigate popularity bias, merely adding the word diversity to LLM prompts fails to convey this semantic intent.
For both Qwen-max and Llama, the Diversity baseline consistently underperforms the Item-rank baseline across key metrics such as LtC, MRR, and F1.
For example, Qwen-max’s MRR decreases from $0.117$ to $0.068$, and its LtC drops from $0.045$ to $0.024$ when diversity is included.
These results indicate that LLMs interpret diversity at a superficial level, selecting less relevant items rather than genuinely mitigating popularity dominance, thereby degrading both accuracy and coverage.
In contrast, providing explicit item-rank information to RecLLM achieves better debiasing performance than the diversity-enhanced prompt.
Although the Diversity prompt occasionally improves MRMC, this gain comes at the cost of recommendation relevance, revealing that LLMs lack an intrinsic semantic understanding of fairness-related concepts such as diversity.
Therefore, effective debiasing in RecLLMs requires structured, context-aware guidance rather than generic textual cues.
This finding is further supported by FairLRM, which incorporates explicit data distribution information and achieves consistently superior debiasing results.

\section{Conclusion}
\par This paper explores the semantic understanding of popularity bias with the emerging RecLLM paradigm, revealing how RecLLMs, despite their flexibility, often inherit and amplify this challenge from traditional systems.
Empirical observations show that prompting for "diversity" alone is insufficient, as LLMs tend to diversify only among popular items while sacrificing recommendation accuracy. 
To address this limitation, we propose FairLRM, a dual-side debiasing framework that explicitly models popularity bias from both the item and user perspectives.
By categorizing users based on their historical preferences for popular and niche items and embedding this contextual information into LLM prompts, FairLRM enables the model to semantically comprehend and mitigate popularity bias in a more principled way.
Experimental results demonstrate that FairLRM substantially improves LtC and reduces MRMC, ensuring fairer exposure of non-popular items without degrading, and often enhancing, overall accuracy (F1 score).
These findings highlight the importance of incorporating user-side semantics into debiasing strategies to move beyond surface-level diversity cues.
In summary, FairLRM provides a practical and effective pathway toward semantically grounded, fair, and accurate recommendation systems powered by LLMs.
Future work will further extend this semantic understanding framework to other bias forms, such as exposure imbalance and group-level preference skew.

\bibliographystyle{ACM-Reference-Format}
\bibliography{fairlrm}

\appendix
\section{Ethical Use of Data and Informed Consent}
All datasets used in this study (e.g., MovieLens-$20$M and Goodbooks-$10$k) are publicly available under their respective research licenses. 
These datasets were released for research purposes and have been widely adopted in prior work on recommender systems and fairness-aware machine learning. 
They do not contain personally identifiable information beyond what has been made publicly accessible, and no additional data collection or annotation was conducted by the authors. 
As this work relies solely on secondary analysis of existing open datasets, no direct interaction with human participants occurred, and informed consent was not required. 
The study complies with the ethical use of data guidelines established by the research community.
\end{document}